\documentclass[aps,prb,twocolumn,groupedaddress]{revtex4}
\usepackage[dvips]{graphics}
\usepackage{vmargin}

\setpapersize{USletter}
\setmarginsrb{1in}{1in}{1in}{1in}{0in}{0.2in}{0in}{0in}

\begin{document}

\title{Quantum interference effects in particle transport through square 
lattices}

\author{E. Cuansing}
\author{H. Nakanishi}
\affiliation{Department of Physics, Purdue University, West Lafayette, IN
        47907}
\date{\today}

\begin{abstract}
We study the transport of a quantum particle through square lattices of
various sizes by employing the tight-binding Hamiltonian from quantum 
percolation.  Input and output semi-infinite chains are attached to the 
lattice either by diagonal point to point contacts or by a busbar connection.
We find resonant transmission and reflection occuring whenever the incident 
particle's energy is near an eigenvalue of the lattice alone (i.e., the
lattice without the chains attached).  We also find the transmission to be 
strongly dependent on the way the chains are attached to the lattice.
\end{abstract}

\pacs{}

\maketitle

Quantum interference effects are important in the transport of particles
in mesoscopic systems.  Consider, for example, a particle traversing 
through a square array of quantum dots.  Assume the distance between dots 
is close enough so that the particle can hop between nearest neighbor 
dots.  Considering only the effect of quantum interference, will the 
particle go through the lattice?  Classically, the particle has a 
multitude of paths to go from one end of the lattice to the other, 
depending on the size of the lattice.  Quantum mechanically, however, 
constructive or destructive interference can occur because of the different 
path lengths.  Thus, the transmission of a particle is not assured even 
when there are classically well-defined paths for it to go through the 
lattice.  In this work we investigate the effects of quantum interference 
in the transport of a particle in discrete and finite square lattices.

We consider the particle to be governed by the tight binding Hamiltonian
from quantum percolation \cite{kirkpatrick:72,degennes:59}.  This Hamiltonian 
has the form
\begin{equation}
  H = \sum_{\left< ij \right>} v_{ij} \left( \left| i \right> \left< j \right|
  + \left| j \right> \left< i \right| \right) ,
\label{eq:Hamiltonian}
\end{equation}
where $\left| i \right>$ and $\left| j \right>$ represent tight binding
basis functions centered on sites $i$ and $j$, respectively, and 
$v_{ij} = 1$ if $i$ and $j$ are nearest-neighbors and $v_{ij} = 0$ 
otherwise.  The sum is only over nearest-neighbors.  In quantum percolation
the particle is confined to traverse through disordered clusters constructed
from the methods of percolation theory \cite{stauffer:94} with some occupation 
probability $p$.  For $p < 1$ there is disagreement whether particle states
are localized or extended.  In a review by Mookerjee, et. al. 
\cite{mookerjee:95}, they concluded that all states are localized and 
transport is dominated by statistically exceptional necklace-like resonant 
states.  Daboul, et. al. \cite{daboul:00}, by calculating the moments of 
distances between pairs of lattice sites using series expansion methods, 
found evidence of a transition from exponentially localized to extended or 
power-law decaying states with an energy-dependent occupation probability 
threshold $p(E)$.  Recent numerical studies of the scaling of the conductance
$g$ by Ha{\l}da{\'{s}}, et. al. \cite{haldas:02}, however, found all states 
to be localized and no indication of a localization-delocalization transition.
In this work we only consider the limiting case $p = 1$ wherein all sites in 
the lattice are occupied, i.e., all sites are available to the particle 
through nearest-neighbor hops, and all particle states are thought to be 
extended.  However, we will show that even in this limit, the transport
{\em through} the lattice is very sensitive to the incident particle's 
energy, varying from complete transmission to complete reflection.

To determine the transport properties of a particle traversing through the
square lattice, we attach semi-infinite chains to the left and right sides 
of that lattice.  Call the left semi-infinite chain the input chain and 
the right semi-infinite chain the output chain.  The particle is made 
incident to the lattice via the input chain.  If the particle goes through 
the lattice, then it exits via the output chain.  Following the 
Landauer-B\"{u}ttiker formalism \cite{buttiker:85}, the conductance of the 
system can then be determined from the resulting transmission and reflection 
amplitudes.  Because of the semi-infinite chains the corresponding matrix 
equation resulting from Eq.~(\ref{eq:Hamiltonian}) is also infinite.  Daboul, 
et. al. \cite{daboul:00} recently described a method to transform the 
infinitely-sized Hamiltonian matrix in Eq.~(\ref{eq:Hamiltonian}) into a 
reduced matrix, $H'$, that is finite and involves only the lattice and its 
connections to the semi-infinite chains using an ansatz.  We are implementing 
this method in this work.

\begin{figure}[h]
{\resizebox{2.5in}{2in}{\includegraphics{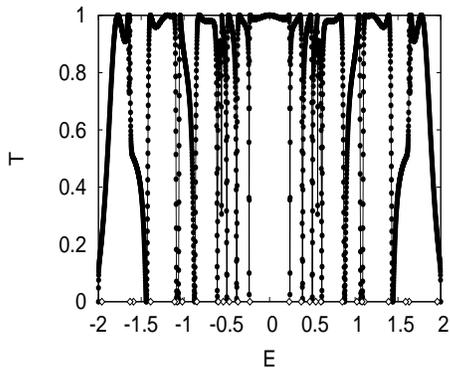}}}
\caption{Plot of the transmission coefficient, $T$, against the incident 
particle's energy, $E$, for a $10 \times 10$ lattice with point to point 
contacts to the input and output chains.  The diamonds ($\diamond$) are the 
locations of the doubly degenerate eigenvalues of the isolated square 
lattice.}
\label{fig:pt2pt}
\end{figure}

There are various ways of attaching the semi-infinite chains to the square
lattice.  In this work we consider two ways.  One is by point to point 
contacts and the other is by a busbar connection.  In point to point 
contacts the input chain is singly attached to the top-leftmost site 
while the output chain is singly attached to the bottom-rightmost site of 
the square lattice.  In a busbar connection the input chain is attached 
to all the sites in the left side of the lattice while the output chain is 
attached to all the sites in the right side of the lattice.

Daboul, et. al. \cite{daboul:00}, proposed the following ansatz:
\begin{equation}
  \begin{array}{ccl}
    \psi_{-(n+1)} & = & e^{-inq} + r e^{inq},\\
    \psi_{+(n+1)} & = & t e^{inq},\\
  \end{array}
\label{eq:ansatz}
\end{equation}
where $n = 0, 1, 2, \ldots$.  The $\psi_{-(n+1)}$ represent components of 
the wavefunction along the input chain and the $\psi_{+(n+1)}$ represent 
components along the output chain.  $\psi_{-1}$ and $\psi_{+1}$
are for the sites in the input and output chain, respectively, that are
directly connected to the lattice.  The ansatz restricts solutions to
Eq.~(\ref{eq:Hamiltonian}) in the form of incident and reflected plane waves
along the input chain and transmitted plane waves along the output chain.
Because of this ansatz the energy of the incident particle is also 
restricted to be within $E = -2$ and $E = 2$.  The transmission and
reflection coefficients can be determined from the $t$ and $r$ in 
Eq.~(\ref{eq:ansatz}) by $T = |t|^2$ and $R = |r|^2$.

\begin{figure}[h]
{\resizebox{2.7in}{2in}{\includegraphics{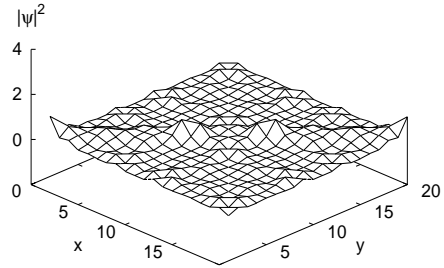}}}\\
\centering{(a)}\\
{\resizebox{2.7in}{2in}{\includegraphics{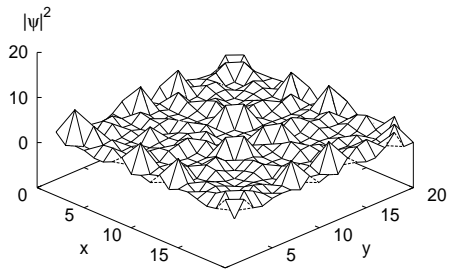}}}\\
\centering{(b)}\\
\caption{Sample states of a particle traversing through a $20 \times 20$ 
lattice with point to point contacts to the input and output chains.  
(a) Highly transmitting state with $E = 0.39$.  (b) Highly reflecting state 
with $E = 0.41$.}
\label{fig:pt2ptstates}
\end{figure}

Once the Hamiltonian matrix in Eq.~(\ref{eq:Hamiltonian}) is reduced to
$H'$, the resulting problem can then be cast into the form of a linear
equation $(H' - E)\psi = \gamma$, where $\gamma$ is solely a function of $E$.  
This linear equation can then be solved for $\psi$ once $E$ is chosen.  We
determine $t$ and $r$ from $\psi$ by numerically solving the above linear
equation exactly, i.e., from $\psi = (H' - E)^{-1} \gamma$.  The matrix 
$(H' - E)$ is sparse and is numerically very close to being singular,
making the use of standard methods such as LU decomposition fail in some
instances.  As such, we implement the technique called singular value
decomposition \cite{press:92} to carefully determine the inverse of 
$(H' - E)$.

Shown in Fig.~\ref{fig:pt2pt} is the plot for the transmission coefficient 
against the incident particle's energy for a $10 \times 10$ lattice with 
point to point contacts to the input and output chains.  Also shown are 
the locations of the doubly degenerate eigenvalues of the isolated square 
lattice.  An isolated lattice is one where the input and output chains are
not attached.  The system is highly transmitting except at some values of 
energy where there are sharp dips and the system becomes highly reflecting.  
Notice that the dips occur near the eigenvalues of the isolated lattice.  
This phenomenon is analogous to the resonant tunneling \cite{datta:95} of an 
incident particle through, for example, a double-barrier system.  In this 
work, however, although there is no tunneling involved, we do see resonant 
reflection whenever the energy of the incident particle falls near a doubly 
degenerate eigenvalue of the isolated lattice.

Notice as well that there is symmetry between the $E > 0$ side and the 
$E < 0$ side.  The square lattice has bipartite symmetry and point to point 
contacts connections preserve that symmetry.  Maintaining bipartite symmetry
can in turn be shown to ensure the symmetry in $T$ about $E = 0$.

\begin{figure}[h]
{\resizebox{2.5in}{2in}{\includegraphics{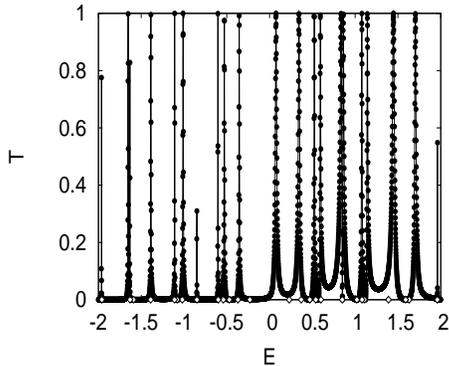}}}
\caption{Plot of $T$ against $E$ for a $10 \times 10$ lattice with busbar 
connections to the input and output chains.  The diamonds ($\diamond$) 
are again the locations of the doubly degenerate eigenvalues of the 
isolated square lattice.}
\label{fig:bsbar}
\end{figure}

As the size of the isolated lattice is increased the number of its 
associated eigenvalues will also increase.  For the lattice with point 
to point contacts to the chains, we also see more dips in the transmission 
coefficient as we increase the size of the lattice.  These dips are also 
located near the doubly degenerate eigenvalues of the corresponding 
isolated square lattice.

Shown in Fig.~\ref{fig:pt2ptstates}(a) and (b) are sample states that are
highly transmitting and highly reflecting, respectively, for a particle 
traversing through a $20 \times 20$ lattice.  The lattice is at the $xy$ 
plane.  The input chain is attached to the site in the lattice located
at $(1,1)$.  The output chain is attached to the site in the lattice 
located at $(20,20)$.  The $z$-axis is the absolute square of the components 
of the wavefunction at each corresponding lattice site, 
$\left| \psi(x,y) \right|^2$.  For the highly transmitting state we see 
a diagonal line of non-zero $\psi$ going from the input to the output 
chains.  Though this is not always true for all highly transmitting
states, those with this feature are always highly transmitting.  In the
highly reflecting state, on the other hand, we see large fluctuations
and destructive interference is manifest at the input and output sites.

\begin{figure}[h]
{\resizebox{2.7in}{2in}{\includegraphics{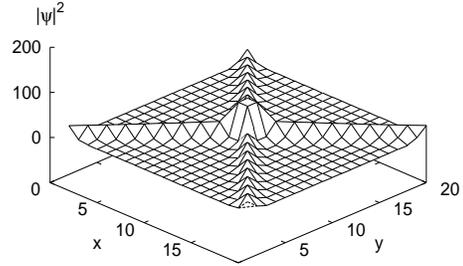}}}\\
\centering{(a)}\\
{\resizebox{2.7in}{2in}{\includegraphics{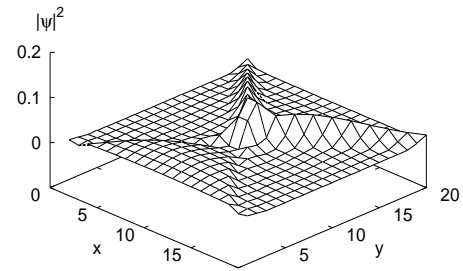}}}\\
\centering{(b)}\\
\caption{Sample states of a particle traversing through a $20 \times 20$ 
lattice with busbar connections.  (a) Highly transmitting state with 
$E = 0.017$. (b) Highly reflecting state with $E = 0.051$.}
\label{fig:bsbarstates}
\end{figure}

Shown in Fig.~\ref{fig:bsbar} is the transmission $T$ versus the incident 
particle's energy $E$ plot for a $10 \times 10$ lattice with busbar 
connections to the input and output chains.  In contrast to the case for
point to point contacts, the system is mostly reflecting but with sharp 
peaks in transmission at certain values of the incident particle's energy.  
Also shown in Fig.~\ref{fig:bsbar} are the locations, as diamonds, of the 
doubly degenerate eigenvalues of the corresponding isolated $10 \times 10$ 
square lattice.  Notice that for the $E < 0$ side the locations of the 
transmission peaks are near the eigenvalues of the isolated lattice.  This 
is similar to the case with point to point contacts but instead of resonant
reflection we see resonant transmission.  For the $E > 0$ side, however, 
some of the peaks do not coincide with the locations of the eigenvalues of 
the isolated lattice.  

From the ansatz shown in Eq.~(\ref{eq:ansatz}), the wave vector $q$ of 
the particle is related to its energy by $E = 2 \cos(q)$, where 
$q = 2 \pi/\lambda$.  For negative energies, the particle's wavelength is 
constrained to be within $\frac{4}{3} < \lambda < 4$.  For positive energies, 
the wavelength should be within either $\lambda < \frac{4}{3}$ or 
$\lambda > 4$.  Unlike the case for point to point contacts, the lack of 
symmetry between the $E < 0$ and $E > 0$ sides of the plot in 
Fig.~\ref{fig:bsbar} indicates the significance of the incident particle's 
wavelength when undergoing through a busbar connection.  Mathematically, the 
multiple connections of the busbar destroys the bipartite symmetry of the 
square lattice, and consequently destroying the symmetry in $T$ about 
$E = 0$.

Let us call those sites at the sides of the lattice that are directly 
connected to the input and output chains as belonging to the input and output 
connection boundaries, respectively.  Because of the multiple connections in 
a busbar, destructive interference can occur at the connection boundaries 
resulting in a vanishingly small transmission through the lattice.
Some of the minima in transmission in Fig.~\ref{fig:bsbar} appear to
be consistent with rules analogous to optical interference minima/maxima
conditions on the boundary.  For example, the condition that an integer
number of wavelengths fit within the boundary of a lattice of size
$L \times L$, i.e., the condition $L - 1 = n \lambda$, would suggest that 
certain values of $\lambda$ result in destructive interference.  This would 
include $\lambda = 1$, i.e., $E = 2$ from $E = 2 \cos(2 \pi/\lambda)$, for 
all $L$, and $\lambda = 2$ ($E = -2$) for all odd values of $L$.  In 
actuality, completely destructive interference occurs when $E = 2$ for all 
$L > 2$ and also when $E = -2$ for all $L \neq 2~{\rm and}~4$.  There are 
also several other minima in $T$ that are consistent with this condition.  
For example, for $L = 5$, $\lambda = 4$ ($E = 0$) also satisfies the 
condition and indeed it is close to a transmission minimum.  For $L = 6$, 
$\lambda = 5/n$ ($n = 1, 2, 3, 4$), corresponding to 
$E \approx -1.62~{\rm and}~0.62$, also satisfy the condition and they are 
also close to a minima of $T$.  In addition, at $\lambda = 2$ ($E = -2$)
we actually observe completely constructive interference for 
$L = 2~{\rm and}~4$, where $L - 1 = (1/2) \lambda$ and 
$L - 1 = (3/2) \lambda$, respectively, are satisfied.  These observations 
suggest strong influences of interference on or near 
the connection boundaries on the overall transmission regarding the busbar 
connection though this boundary interference effect is far from providing a 
satisfactory explanation.  In fact, since we have a discrete system with 
unit lattice constant rather than a continuous {\em slit} as in an optical 
system, it is not clear why $\lambda = 1$ actually leads to destructive 
interference rather than the opposite (except for $L = 2$).  Of course, 
any influence of interference along the connection boundary must only be 
a part of the story since interference actually occurs throughout the bulk 
of the system (on most of which $\lambda$ is not even well-defined) and 
since it must also compete with resonant transmission and reflection whenever 
the values of the incident particle's energies at the input chain fall near 
the eigenvalues of the isolated cluster.

Two sample states for a particle traversing through a $20 \times 20$ lattice 
with busbar connections to the input and output chains are shown in 
Fig.~\ref{fig:bsbarstates}.  The busbars are connected at the $y = 1$ and 
$y = 20$ sides of the lattice.  Shown in Fig.~\ref{fig:bsbarstates}(a) is 
a highly transmitting state while Fig.~\ref{fig:bsbarstates}(b) is a highly 
reflecting state.  Notice that the difference in amplitudes between the 
states is several orders of magnitudes.  In Fig.~\ref{fig:bsbarstates}(b) 
strong destructive interference occurs in such a way that the state $\psi$ 
nearly vanishes within the lattice.

In conclusion, we find resonant transmission and reflection in the
transport of a particle through finite square lattices whenever the 
particle's energy is near an eigenvalue of the isolated lattice.  The 
way the input and output chains are attached to the lattice influences
the transport behavior of the incident particle.  For point to point
contacts the particle is mostly transmitting but with transmission dips
whenever resonance occurs.  For busbar connections the particle is
mostly reflecting with transmission peaks whenever resonance also
occurs.  There are, however, peaks in transmission that can not be
accounted for by resonance.  These peaks are results of interference
originating from the multiple connections in a busbar.

\vspace{0.1in}

We would like to thank Y. Goldschmidt, Y. Lyanda-Geller, G. Baskaran, 
A. Finkelstein, L. Rokhinson, G. Giuliani and N. Giordano for fruitful
discussions.

\bibliography{references}

\end{document}